\documentclass[conference]{IEEEtran}
\IEEEoverridecommandlockouts
\usepackage{cite}
\usepackage{amsmath,amssymb,amsfonts}
\usepackage{algorithmic}
\usepackage{graphicx}
\usepackage{textcomp}
\usepackage{xcolor}
\usepackage{balance}
\def\BibTeX{{\rm B\kern-.05em{\sc i\kern-.025em b}\kern-.08em
    T\kern-.1667em\lower.7ex\hbox{E}\kern-.125emX}}
\begin{document}
\pagenumbering{gobble}

\title{\textbf{\Large Importance of Coordination and Cultural Diversity for an Efficient and Flexible Manufacturing System}}

\author{%
Kashif Zia$^{1}$, Alois Ferscha$^{2}$ and Dari Trendafilov$^{3}$
\thanks{$^{1}$K. Zia is with the Faculty of Computing and Information Technology, Sohar University, Oman. He is also adjunct senior researcher with the Institute of Pervasive Computing, Johannes Kepler University, Linz, Austria.
        {\tt\small kzia@su.edu.om}}%
\thanks{$^{2}$A. Ferscha is with the Institute of Pervasive Computing, Johannes Kepler University, Linz, Austria.
        {\tt\small ferscha@pervasive.jku.at}}%
\thanks{$^{3}$D. Trendafilov is with the Institute of Pervasive Computing, Johannes Kepler University, Linz, Austria.
        {\tt\small dari.trendafilov@pervasive.jku.at}}%
}

\author{\IEEEauthorblockN{Kashif Zia}
\IEEEauthorblockA{\textit{Faculty of Computing and IT} \\
\textit{Sohar University, Oman}\\
Email: kzia@su.edu.om}
\and
\IEEEauthorblockN{Alois Ferscha}
\IEEEauthorblockA{\textit{Institute of Pervasive Computing} \\
\textit{Johannes Kepler University, Austria} \\
Email: ferscha@pervasive.jku.at}
\and
\IEEEauthorblockN{Dari Trendafilov}
\IEEEauthorblockA{\textit{Institute of Pervasive Computing} \\
\textit{Johannes Kepler University, Austria} \\
Email: dari.trendafilov@pervasive.jku.at}
}

\maketitle
\begin{abstract}
Manufacturing systems of the future need to have flexible resources and flexible routing to produce extremely personalized products, even of lot size equal to one. In this paper, we have proposed a framework, which is designed to achieve this goal. Towards this, we have integrated an established cultural evolution model to achieve desired flexibility of resources and acceptable routing time. Promising results are evidenced through a simple proof-of-concept simulation.
\end{abstract}


\IEEEpeerreviewmaketitle

\section {Introduction}
The industrial manufacturing paradigm has already evolved from mass production to mass customization. Fueled by initiatives like Industry 4.0 \cite{lee2015cyber}, we foresee a further improvement in the coming years, namely the paradigm of personalized production. Personalized production targets an extremely flexible manufacturing system which could respond to predicted and unpredicted changes in the production environment.  

\vspace{-1mm}
According to \cite{ogunsakin2018bee}, this flexibility should be a collection of three aspects, at least:
\vspace{-1mm}
\begin{itemize}
    \item Resource Flexibility: flexibility of machines/processing stations to make multiple parts.
    \item Routing Flexibility: flexibility to execute the same operation/function using multiple processing stations.
    \item Lot Size Flexibility: ability to produce a very small customized/personalized lot size in a non-batch mode.
\end{itemize}
\vspace{-1mm}
Historically, many research efforts have focused on specific features of these aspects. Many scheduling \cite {wang2016mpn} \cite{marichelvam2014discrete}, resource optimization \cite{ogunsakin2018bee} \cite{beruvides2018artificial}, and constraint satisfaction \cite{ezpeleta1995petri} solutions have been presented. However, all these mechanisms either consider a mathematical abstraction or imitate a real-world situation as their manufacturing environment. The problem is that this results in a static configuration, and the solution proposed only works in these boundaries. 

\vspace{-1mm}
For modeling of a dynamical system, it is imperative to use a computational approach. For example, a more recent work uses an agent-based model while considering mobile processing stations as a mean to achieve flexibility in the manufacturing process \cite{ogunsakin2018bee}. The idea is to make resources available when and where these are required. Although their approach addresses the challenge of routing flexibility to an extent, the capabilities of resources still remain static. 

\vspace{-1mm}
In our research, we are mostly focusing on resource flexibility, which means that the processing units are able to dynamically change their \textit {capabilities} and therefore a resource is able to perform several tasks. The goal is to keep resources stationary (and avoid expensive process of mobility) and arrange resources in groups of \textit {complementing} capabilities. Ideally, a resource would designate itself for a capability that would optimize the manufacturing process in several dimensions, such as production rate, lead-time per order and reactivity index \cite{ogunsakin2018bee}.    

\vspace{-1mm}
In this context, our mechanism exploits the cultural nature of the scenario; i.e. groups of \textit {complementing} capabilities; and we are convinced that cultural diversity (complementation) has a lot of potentials to explore about. We argue that flexibility in resources, routing and personalizing closely relate to the evolution of culture, particularly cultural groups and diversification. This would provide an entirely new perspective for future research in this domain. 

\vspace{-1mm}
A culture is a multi-featured system evolving in time. One characteristic of culture is its coherence when seen from outside. Definitely, this coherence results due to a majority of people trying to acquire a similar behavior (often termed as a trait) in a certain context (often termed as a feature). Hence, a conceptual framework comprising of resources and products, driven by related features and traits can be formulated. A resource is a processing unit in the production line, whereas a product is obviously a product under production. Although a product can also be considered as a cultural entity, it is not the case for now. Only a resource is a cultural entity. The framework particularly focuses on limited coherence between cultural groups.

\vspace{-1mm}
Resources are flexible, initially having some randomly chosen features and a randomly chosen trait against a feature. For example, a processing unit may have ability to perform one, two or more tasks {$T_1$, $T_2$, ...} with certain levels of precision {$P_1$, $P_2$, ...}. Here, a tuple consisting of $n$ values is a set describing capabilities of a resource. For example, the set \{$P_2$, $P_1$, $P_3$\} can be interpreted as: this resource can perform task 1 with precision 2, task 2 with precision 1 and task 3 with precision 3. Furthermore, it cannot perform any other task. 

\vspace{-1mm}
Such a scheme is naturally compatible with the requirement of a flexible manufacturing system stated above, namely, flexibility in resources, routing and personalizing. Axelrod provides evidence in his seminal work \cite{axelrod1997dissemination} for such a simple configuration of cultural descriptions which can result in a locally coherent, but globally polarized culture as a consequence of localized interactions of participating entities.  

\vspace{-1mm}
However, in this paper, we argue that such a limitless coherence has no control over where the boundaries of the global polarization would occur, which turned out to be harmful to a system which usually seeks for the economy of resources and optimizations in several dimensions. That is the reason, we try to find conditions which end up in approximately acceptable structuring in terms of coherence (termed as limited coherence) vs. polarization. To achieve this, we have used and refined Axelrod's model of cultural dissemination \cite{axelrod1997dissemination}.

Axelrod's model provides evidence of observation that the more time we provide for cultural dissemination, the cultural groups become increasingly coherent due to homophily. For scenarios, which require diversification of resources, we need to find a balance between coherence and diversification. This paper provides first insights into these aspects for a production shop floor. The paper presents an agent-based model, abstracting and simplifying the production process at a hypothetical shop floor. 

\vspace{-1mm}
The rest of the paper is structured as follows. In section \ref{sec:methods}, a detailed description of the methods of modeling and simulation is given, followed by a discussion on initial findings in section \ref{sec:findings}. We end the paper with an elaborate outlook of future work given in section \ref{sec:outlook}.

\vspace{-2mm}

\section {Methods} \label{sec:methods}

In the following, a detailed description of the models is given. Starting from desperation of Axelrod's model of cultural dissemination, next the motivation of the proposed model is given, followed by the details of the proposed method itself.   

\subsection {Axelrod's Model of Cultural Dissemination}

\vspace{-1mm}

Axelrod's model \cite{axelrod1997dissemination} thrived for cultural homogeneity \cite{bednar2010emergent}, where adjacent cultures gets influence from each other. The model is based on cultural components defined by three factors; features, traits, and persons. Culture has many features, such as habits of eating, recreation, and leisure. These features may not be identical across different cultures. Each of these features has several traits, which may differ across cultures. A person is a placeholder of a culture described by one of $f$ features and $t$ traits. Axelrod proposed a model seeking for cultural homogeneity proclaiming that different cultures are destined to cohere together so that they appear as a cultural unity, but at the same time, there exists a clear-cut differentiation between cultures.

\vspace{-1mm}

\begin{figure}
\centering
\includegraphics[width=0.40\textwidth]{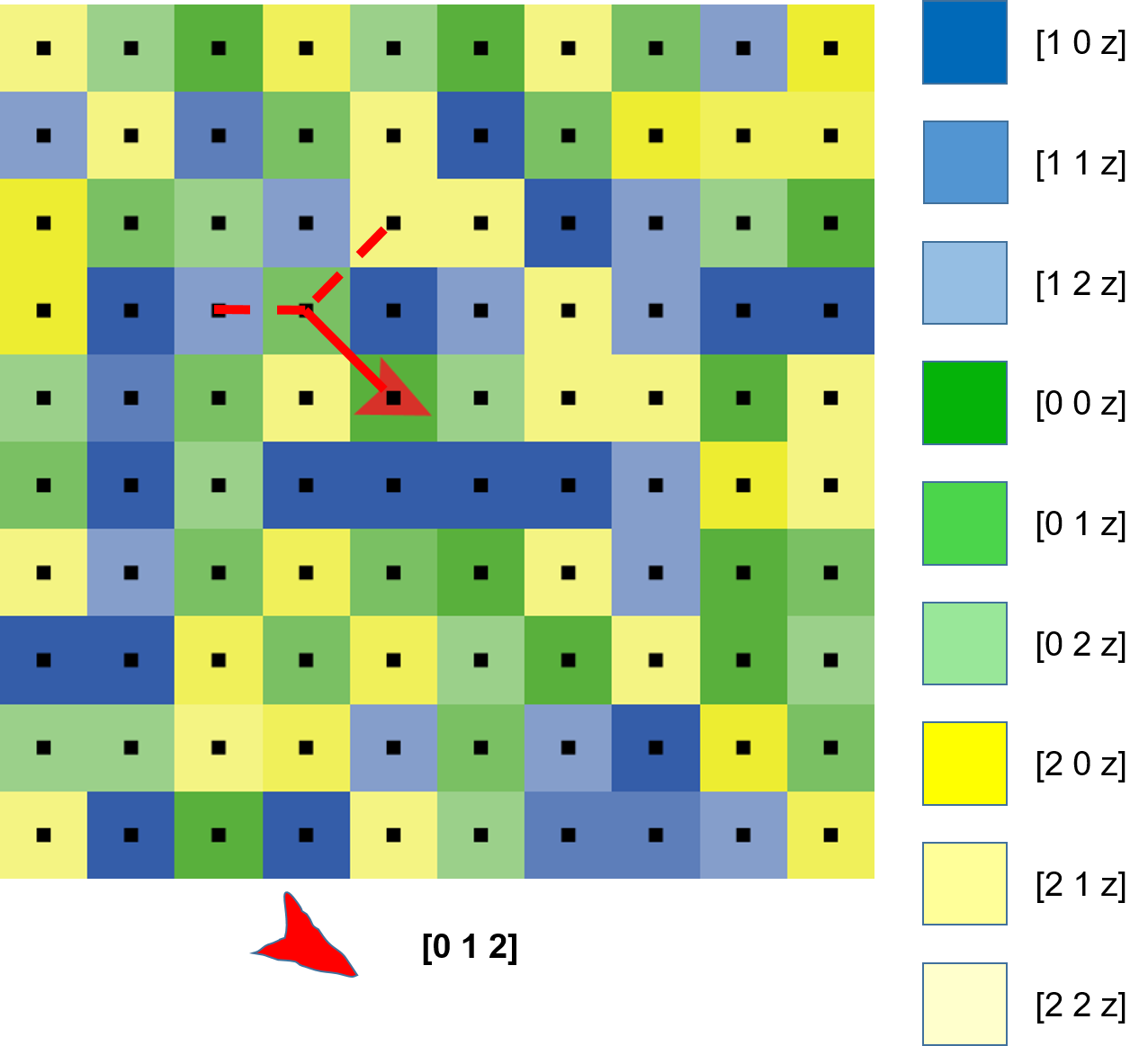}
\caption{Initial distribution of a $10 \times 10$ grid constituted by blocks of culture; each block a tuple of 3, representing three features (green, blue, yellow) having three traits (3 shades of a color) each. Average $diversity\_index$ of random setting is around 0.660.}
\label{fig:init}
\end{figure}

\vspace{-1mm}

Axelrod model was able to demonstrate that the above two (rather contradictory) goals can be achieved by a simple interaction model (realized through N coordination games) between neighboring persons. Axelrod showed that N coordination games are necessary for a broader scale evolution of culture. Furthermore, groups' consistency across different aspects of societal norms makes a group culturally coherent and different from others. In the following, a hypothetical case study representing Axelrod's model of cultural diversification is presented. 

\begin{figure}[h]
\centering
\includegraphics[width=0.49\textwidth]{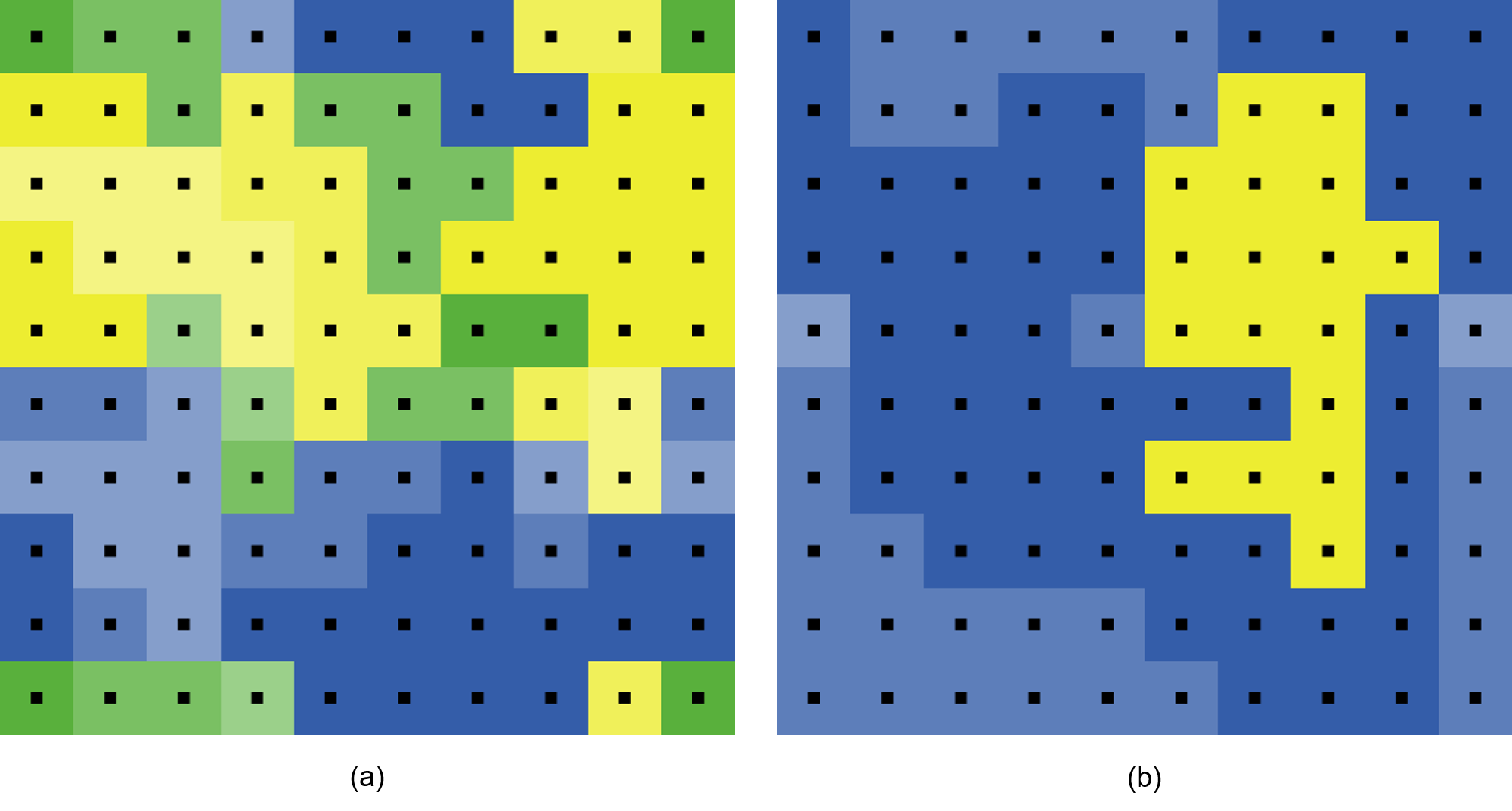}
\caption{Axelrod's Model: Evolution of cultures shown in Figure \ref{fig:init}. (a) at simulation iteration 6055 showing clusters of cultures starting to form. (b) at simulation iteration 12207 showing further consolidation of clusters of cultures. The evolution is destined to end up in very few cultures (1 or 2).}
\label{fig:AxelrodClusters}
\end{figure}

In Figure \ref{fig:init}, a grid of $10 \times 10$ cells is shown. Each cell is represented by a person or a culture depicted by color (one unique combination out of $f \times t$ possible combinations). Each cell's color has a meaning; for example, all green cells have the capability to perform task 1 with precision value 0, which is followed by precision values of task 2 (0, 1 or 2); last value is not path dependent and represented by z. A product has a unique sequence of the task to perform represented with an arrow shape (at the center of the space).

Axelrod model calculated similarity $s$ between neighboring cultures. If $s$ is not 1 ($100\%$), with a probability $p$, the value of a $different$ column of a person is replaced by the corresponding value of the neighboring person. This simple mechanism is able to generate clusters of coherent cultures as shown in Figure \ref{fig:AxelrodClusters}.     
If we define \textbf {diversity\_index} as the mean diversification of cultures of all persons when compared to their neighbors, the Axelrod model would converge into a single culture most of the time with $diversity\_index$ equal to 0. This is not desirable in the context in which we want to use this model. Therefore, the model was extended as detailed in the following.

\vspace{-1mm}

\subsection {The Motivation: Constrained, N-Coordination Games for Cultural Diversity}

\begin{figure}[h]
\centering
\includegraphics[width=0.49\textwidth]{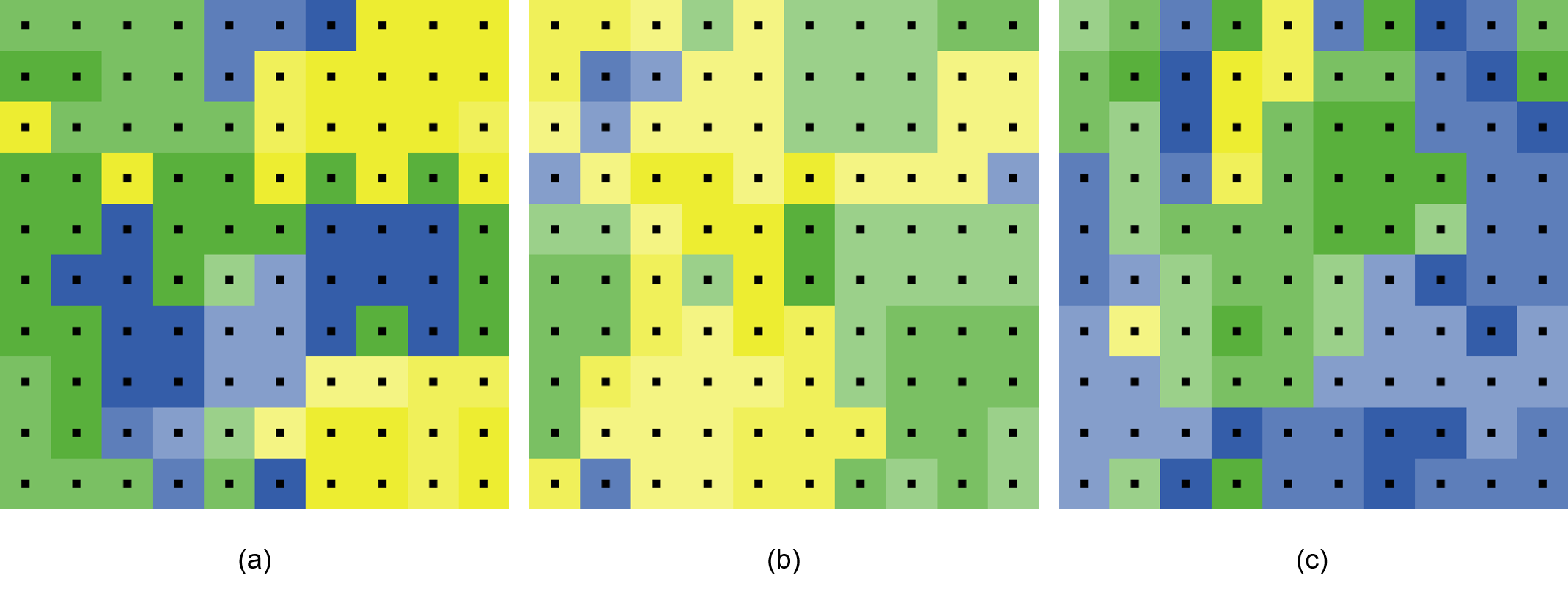}
\caption{Extended Axelrod's Model: Cultural Diversity at iteration 50000. Three random outcomes shown in (a), (b) and (c) having a $diversity\_index$ of around 0.320.}
\label{fig:mymodel50000}
\end{figure}


Before the description of the model, we formalize the scenario given in Figure \ref{fig:init} as a manufacturing process. Give that a processing unit is able to perform three possible tasks with three possible precision values, we can see a clear capability matching through colors. Further, a product is introduced which need to complete a sequence of three tasks offered by different resources. We hypothesize that using the constraint, N coordination games, we can achieve cultural diversity, which is closer to what is desirable. This would directly impact products' traversing efforts in a positive way. A comparison of Figure \ref{fig:AxelrodClusters} with Figure \ref{fig:mymodel50000} shows less diversification from prior to the later. We hypothesize that this would help in reducing the traversing efforts of the products.   

\vspace{-1mm}

\noindent An Example Walkthrough: Referring to Figure \ref{fig:init} again, each resource (black agent at the center of a cell) is randomly populated 
with vector [x y z], where x, y, and z may have three possible values 0, 1 and 2.
The product has to perform three tasks in a sequence. 
Task 1 with precision 0, task 2 with 
precision 1 and task 3 with precision 2.
It starts at the shown position. First, it will perform task 1 with precision 0. That is right away available at the cell the product is situated. 
Next, it has to perform task 2 with precision 1. The nearest resource, which has first column equal to 0 (assuming a connection between task 1 and 2) and second column equal to 1 is the resource on immediate top-left; hence the product would move there.
Next task is task 3 with precision 2. Assuming that it is an independent task, the product would try to find the nearest resource that has the third column equal to 2 (any color). This can be any resource (two valid possibilities are shown with dotted lines).

\vspace{-1mm}

It seems that random configurations would be the best, but this cannot be the case in a structured environment, particularly in case of an assembly line type of manufacturing. The Axelrod model is too skewed towards coherence and would end up in too few cultures. Hence we propose to refine the Axelrod model in the following way.

\vspace{-1mm}

\subsection {The Proposed Diversity Dissemination Mechanism}

Axelrod model sought for similarity $s$ between neighboring cultures. If $s$ is not 1 ($100\%$), with a probability $p$, the value of a $different$ column of culture is replaced by the corresponding value of the neighboring culture. We extend this model by applying an extra constraint. That is, the replacement is only possible if $s$ is also less than a threshold $th$, which is for now given a static value of 0.5. This is expected to increase overall \textit{diversity\_index} of the system. Before analyzing the impact of this refinement the mechanism of product traversing is explained.

\vspace{-1mm}
\subsection {Traversing Mechanism}

All products have a sequence of tasks to perform in the form [x, y, z]. A product first gets the value x, and maps it onto resources with an identical capability and residing close to its position. Let's denote the resource at $r$. After visiting $r$, the product seeks for the next nearest resource corresponding to y. It is assumed that y has a relationship with x. This means that, in terms of colors, this cell (and the resource residing on top of it) should have the same color. The last task z is independent and just show the range of flexibility that the system may have. 

\vspace{-1mm}

\section {Analysis of Initial Findings} \label{sec:findings}


\begin{figure}
\centering
\includegraphics[width=0.49\textwidth]{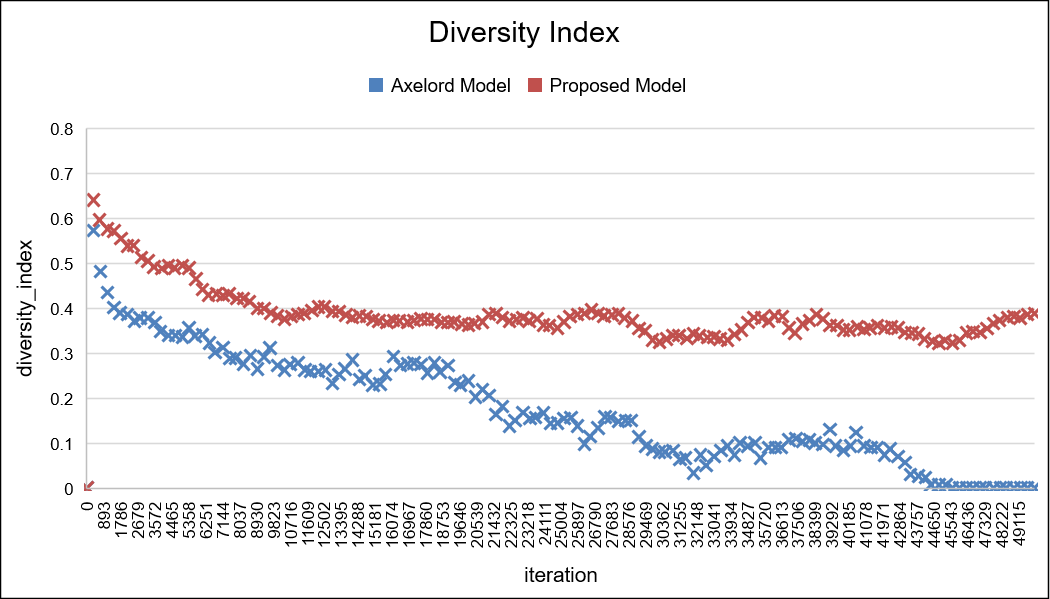}
\caption{Comparison of time series of $diversity\_index$.}
\label{fig:graphNew}
\end{figure}

\begin{figure}
\centering
\includegraphics[width=0.37\textwidth]{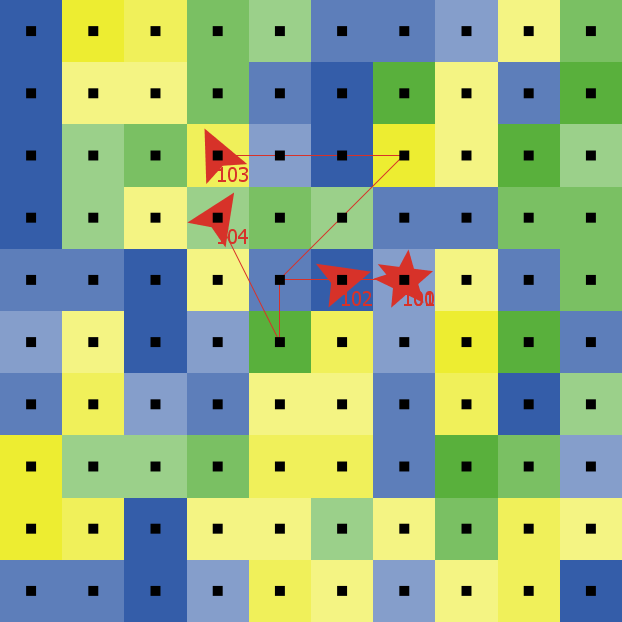}
\caption{Traversing behavior in random configuration of resource capability.}
\label{fig:traverserandom}
\end{figure}

Definitely, the introduction of threshold $th$ retains $diversity\_index$ in case of extension of Axelrod's model, as shown in Figure \ref{fig:graphNew}. This helps in task completion capability of the system due to the provisioning of a more diverse array of complementing capabilities. The graph shown in Figure \ref{fig:graphNew} evidences this fact. The diversity\_index of the proposed model is much higher than the Axelrod's model throughout the simulation and it never dies out no matter how long the system evolves, unlike Axelrod's model.This can be verified by analyzing the mobility of products in case of random configurations. In Figure \ref{fig:traverserandom}, we can see five products which start from the center of the space. Products 100 and 101 needed to perform task 1, so they did it as the first step using the resource where product 102 is stationed now. Next, they move to the right and performed task 2. That means that both completed the first two task of their schedule successfully. The similar is true for the other three products. The system could acquire a mobility index equal to 2 on average for the first two tasks, which it did without any problem. As we mentioned already, a random configuration is most flexible and would always be best in its task completion capability. However, this configuration is unrealistic. In reality, we need to plan the placement of resources and put them in order.  

\vspace{-1mm}

\begin{figure}
\centering
\includegraphics[width=0.49\textwidth]{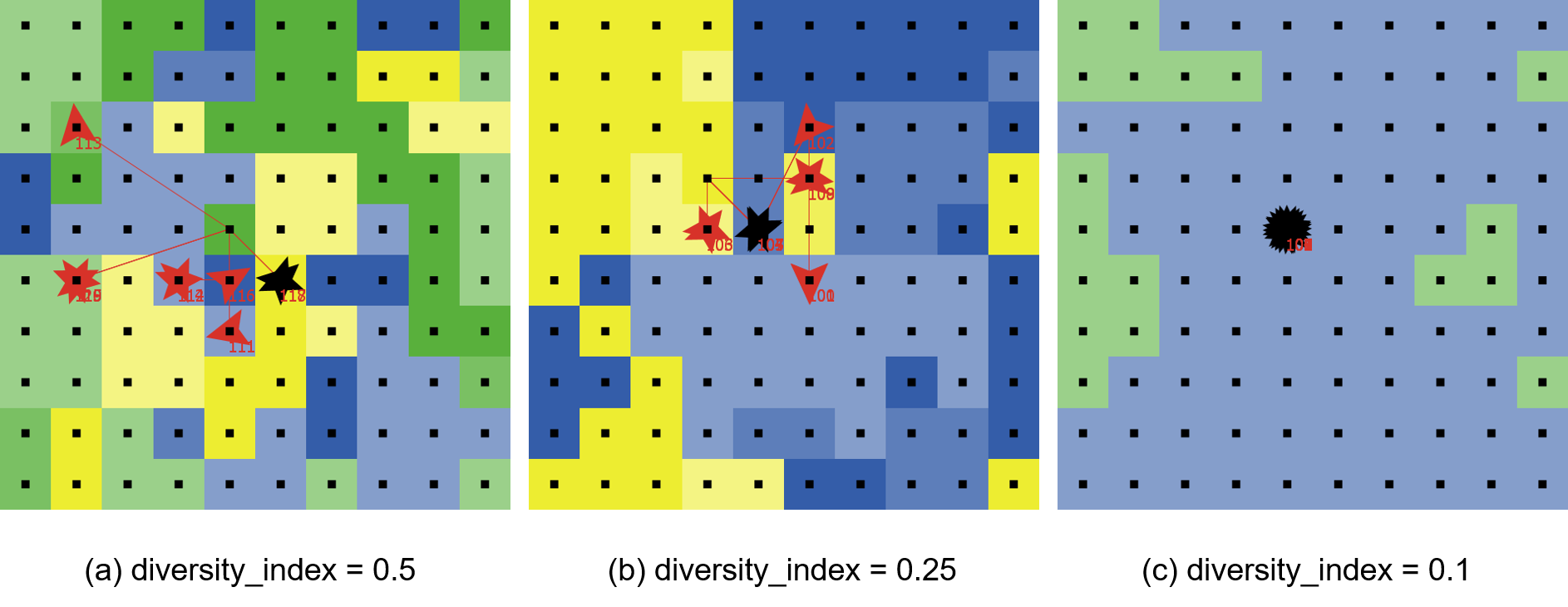}
\caption{Traversing behavior in Axelord's Model.}
\label{fig:traverseaxelrod}
\end{figure}

\vspace{-1mm}

In case of Axelrod's model, we have analyzed the results for $diversity\_index$ 0.5, 0.25 and 0.1. These three situations are represented in Figure \ref{fig:traverseaxelrod}. With increasing polarization and decreasing $diversity\_index$, the average mobility index drops. After running the simulation several times, it was observed that mobility index is 1.8 ($diversity\_index$ = 0.5), 1.4 ($diversity\_index$ = 0.25) and 0.03 ($diversity\_index$ = 0.10). As shown in Figure \ref{fig:traverseaxelrod}, this decrease is due to nonavailability of resources indicated by products turning into black color. 

\vspace{-1mm}

\begin{figure}
\centering
\includegraphics[width=0.49\textwidth]{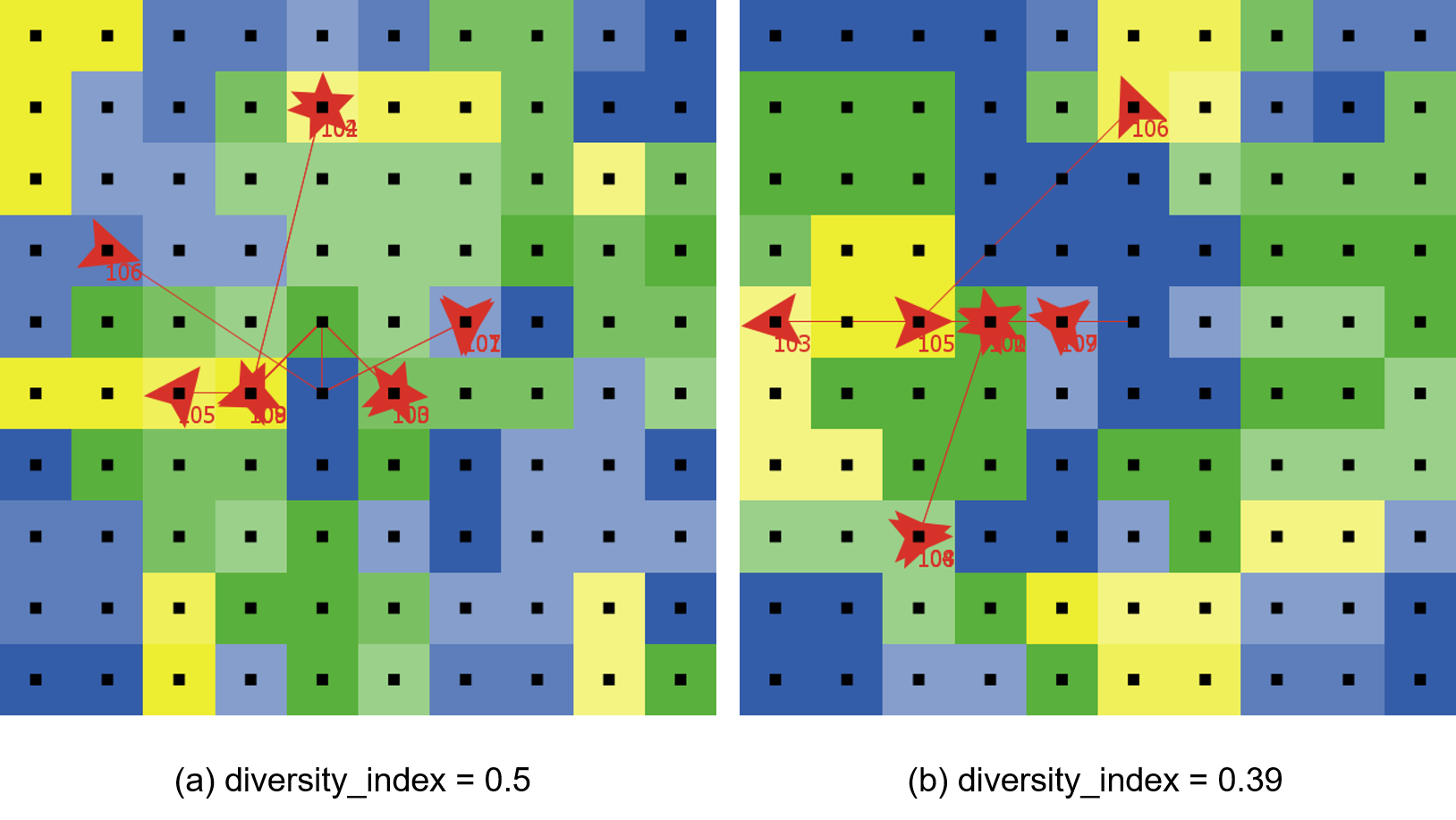}
\caption{Traversing behavior in Extended Axelrod's Model.}
\label{fig:traverseextended}
\end{figure}

\vspace{-1mm}

Lastly, the extended models solve the above issue. We can see a smooth performance of tasks for all the products, which is evident from Figure \ref{fig:traverseextended}. For even minimum possible diversity (0.33), in the majority of the cases, the mobility index achieved is 2 (the possible maximum).

\vspace{-2mm}

\section {Conclusion and Future Work} \label{sec:outlook}

\vspace{-1mm}

Manufacturing systems of the future need to have flexible resources and routing to produce an extremely personalized product, even of lot size equal to one. What we have seen is that flexible manufacturing system can be realized without moving the resources (processing units) by enabling reconfiguration of capabilities of resources based on dissemination of culture concept proposed by Axelrod. However, the Axelrod model has a focus on the coherence of cultural groups, which most of the times end up in one or very few cultures. If we equate such an instance of a culture with a single capability of a resource, we are left with extremely limited resources and products cannot complete their production life cycle.

\vspace{-1mm}

Hence, we proposed to have a constrained cultural coherence mechanism by introducing a threshold. This tiny development has a significant impact on the increase in diversity of the culture along with related resources being in close vicinity to each other on average. This did not only ensure an increase in resource availability as a whole, but also managed to decrease the mobility of products in search of suitable resources.  

\vspace{-1mm}

However, the real contribution of the paper is the integration of manufacturing processes with cultural considerations, which naturally fits into the problem. In our view, this is a novel approach of real significance. However, the work reported in this paper is just a proof-of-concept. We need to have more thorough experiments to measure the efficiency of the model in challenging environments such as environments having inflow and outflow points, more in-depth capabilities and richer relationships between tasks.

\vspace{-1mm}

In the next phase of the project, we will induct models of dynamics, which will include timing of tasks, conflict and deadlock resolution between products seeking identical resources, and more realistic analytics such as production rate, lead-time per order and reactivity index. Lastly, we would also include an autonomous learning system, which would help resources learn and change their configurations on the fly based on product types, requirements, and trajectories.

\vspace{-1mm}

\section*{Acknowledgment}
The authors would like to acknowledge support by FFG funded Pro\(^2\)Future under contract No. 6112792.

\vspace{-1mm}

\bibliographystyle{IEEEtran}
\bibliography{main}

\end{document}